\documentstyle[twoside,fleqn,epscrc2]{article}
\def\lsim{\mathrel{\raise.2ex\hbox{$<$}\hskip-.8em\lower.9ex\hbox{$\sim$}}}
\def\gsim{\mathrel{\raise.2ex\hbox{$>$}\hskip-.8em\lower.9ex\hbox{$\sim$}}}

\newcommand{\AmS}{{\protect\the\textfont2
  A\kern-.1667em\lower.5ex\hbox{M}\kern-.125emS}}

\vspace{-1cm}
\title{QCD Corrections to Jet Cross Sections in DIS\\[-1.8cm]
        \thanks{
         Invited talk presented by E. Mirkes. 
         To appear in proceedings of 
         "QCD and QED in Higher Orders" 1996 
         Zeuthen Workshop on Elementary Particle Theory, April 22-26, 1996.} 
\hfill$\vtop{   \hbox{\normalsize TTP96-24}
                \hbox{\normalsize MADPH-96-946}
                \hbox{\normalsize June 1996}}$ 
        }

\author{
        Erwin Mirkes\address{Institut f\"ur Theoretische Teilchenphysik, 
        Universit\"at Karlsruhe, D-76128 Karlsruhe, Germany   }
        and
        Dieter Zeppenfeld\address{Department of Physics, 
        University of Wisconsin, Madison, WI 53706, USA}}
\begin{document}
\begin{abstract}
Next-to-leading order corrections to
jet cross sections in deep inelastic scattering at HERA are
studied.
The predicted jet rates allow for a precise
determination of $\alpha_s(\mu_R)$ at HERA over a wide range of
$\mu_R$.
We argue, that the ``natural'' renormalization and
factorization scale is set by the average  $k_T^B$
of the jets in the Breit frame and suggest to 
divide the data in corresponding $<k_T^B>$ intervals.
Some implications
for the determination of the gluon density and
the associated forward jet production in the low $x$ regime at HERA
are briefly discussed.
\end{abstract}

\maketitle
\section{Introduction}
Deep inelastic scattering (DIS) at HERA is a copious source of 
multi-jet events. Typical two-jet cross sections
are in the 100~pb to few nb range and thus provide sufficiently high 
statistics for precision QCD tests~\cite{exp_as}.  
Clearly, next-to-leading order (NLO) QCD corrections
are mandatory on the theoretical side for such tests.
Full NLO corrections for one and two-jet production cross sections
and distributions are now available and implemented in the 
$ep \rightarrow n$ jets event generator MEPJET,
which allows to analyze 
arbitrary jet definition schemes and 
general cuts in terms of parton 4-momenta\cite{mz1,mz2}.
A variety of topics can be studied with these tools. 

1) The
determination of  $\alpha_s(\mu_R)$ over a range of scales $\mu_R$
from dijet production at HERA:
The dijet cross section
is proportional to $\alpha_s(\mu_R)$ at leading order (LO), thus suggesting 
a direct measurement of the strong coupling constant. However, the
LO calculation leaves the renormalization scale $\mu_R$ undetermined.
The NLO corrections substantially reduce the renormalization and 
factorization scale dependencies which are present in the LO calculations 
and thus reliable cross section predictions in terms of $\alpha_s(m_Z)$
(for a given set of parton distributions)
are made possible.
Clearly, a careful study of the choice
of scale in the dijet cross section
is needed in order to extract a reliable value for 
$\alpha_s(M_z)$.
We will present some arguments
and  studies of the scale dependence
of the NLO dijet cross section which suggest that the
average $k_T^B$ of the jets in the Breit frame provides
the ``natural'' scale for multi jet production in DIS.
Here, $(k_T^{B}(jet))^2$ is defined by 
$2\,E_j^2(1-\cos\theta_{jP})$, where the subscripts $j$ and $P$
denote the jet and proton, respectively (all quantities are defined
in the Breit frame).

2) The measurement of the 
      gluon density in the proton (via $\gamma g\to q\bar q$):
      The gluon density can only be indirectly constrained
      by an analysis of the structure  function $F_2$ at HERA
      \cite{exp_gluon}.
      The boson gluon fusion subprocess         
      dominates the two jet cross section
      at low $x$ and allows for a more direct measurement of the gluon density
      in this regime. A first LO  experimental analysis
      has been presented in \cite{h1_gluon}.
      NLO corrections reduce the factorization scale dependence
      in the LO calculation (due to the initital state collinear factorization,
      which introduces a mixture of the quark and gluon densities 
      according to the  DGLAP (Altarelli-Parisi) evolution)  
      and thus reliable cross section predictions in terms of the
      scale dependent  parton distributions are made possible.
      Some implications for the determination of the
      gluon density have  been discussed in \cite{mz2}.

3) Associated 
      forward jet production in the low $x$ regime as a signal of BFKL
      dynamics:
      BFKL evolution \cite{bfkl} leads to a larger cross section for 
      events with a measured forward jet (in the proton
      direction) with transverse momentum $p_T^{lab}(j)$ 
      close to $Q$ than the DGLAP \cite{dglap} evolution.
      Clearly, next-to-leading order QCD corrections
      for fixed order QCD, with Altarelli-Parisi (DGLAP) evolution, 
      are mandatory on the theoretical side in order to establish a 
      signal for BFKL evolution in the data.
      We discuss these corrections in section 3.

The importance of higher order corrections
and recombination scheme dependencies 
of the two jet cross sections
for  four different jet algorithms 
(cone, $k_T$ \cite{kt}, JADE, W) was already discussed
in \cite{mz1,mz2}.
While the higher order corrections and recombination scheme
dependencies in the cone
and $k_T$ schemes are small, very large corrections appear in the $W$-scheme. 
We conclude from these studies that the
cone and $k_T$ schemes appear better suited for precision
QCD tests  and will concentrate only on those schemes
in the following.

The goal of a versatile NLO calculation is to allow for an easy 
implementation of an arbitrary jet algorithm and to impose any kinematical
resolution and acceptance cuts on the final state particles. This is best 
achieved by performing all hard phase space integrals numerically, 
with a Monte Carlo integration technique. 
This approach also allows an investigation of the recombination scheme
dependence of the NLO jet cross sections. For dijet production at HERA such
a NLO Monte Carlo program is MEPJET\cite{mz1}.
The calculation is based on the phase space slicing method
and on the technique of universal ``crossing functions'' 
\cite{giele1,rgiele,rk}.
More details are described in \cite{mz1}
and we do not  repeat  them here.
This technique can also be extended to the case of massive
quark production \cite{laenen}.
An alternative technique for the calculation of NLO corrections in
jet physics is the ``subtraction'' method
\cite{catani}.

\section{Choice of the Renormalization and Factorization Scale in 
Multijet Production}
Jet production in DIS is a multi-scale problem
and it is not a priori clear at which scale $\alpha_s$ is probed.
In the following we will discuss the choice of the renormalization and 
factorization scale for dijet production at HERA. 
As mentioned before, the NLO corrections 
are expected to reduce  the scale
dependencies  in the LO predictions
provided the scale is of the order of the  typical hardness
of the partonic process.

The following  studies are done for the
cone algorithm (which is defined in the laboratory frame) and
the distance 
$\Delta R=\sqrt{(\Delta\eta)^2+(\Delta\phi)^2}$ between two partons 
decides whether they should be recombined to a single jet. Here the variables 
are the pseudo-rapidity $\eta$ and the azimuthal angle $\phi$. We 
recombine partons with $\Delta R<1$.
Furthermore, a cut on the jet transverse momenta of $p_T(j)>5$~GeV in the lab
is imposed. We employ the two loop formula for the strong coupling constant
with 
a value for ${\Lambda_{\overline{MS}}^{(4)}}$
consistent with the value from the parton distribution functions.
The value of $\alpha_s$ is matched at the thresholds $\mu_R=m_q$ and the
number of flavors is fixed to $n_f=5$ throughout, {\it i.e.} gluons are 
allowed to split into five flavors of massless quarks.
A running QED fine structure constant $\alpha(Q^2)$ is used.
In addition the following set of kinematical 
cuts is imposed on the initial virtual photon 
and on the final state electron and jets:
We require 
40~GeV$^2<Q^2<2500$ GeV$^2$,
$0.04 < y < 1$, an energy cut of $E(e^\prime)>10$~GeV on the scattered 
electron, and a cut on the pseudo-rapidity $\eta=-\ln\tan(\theta/2)$
of the scattered lepton and jets of $|\eta|<3.5$.

First studies of the scale dependence of the dijet cross section in the cone
scheme are presented in \cite{mz1}.
We have considered scales related to the 
scalar sum of the parton transverse momenta in the Breit frame,
$\sum_i \,p_T^B(i)$, and the virtuality $Q^2$ of the incident photon.
In the following we will also consider scales related to
$\sum_i \,k_T^B(i)$.
Here, $(k_T^{B}(i))^2$ is defined by 
$2\,E_i^2(1-\cos\theta_{iP})$, where the subscripts $i$ and $P$
denote the final parton (or jet) and proton, 
respectively (all quantities are determined
in the Breit frame). 
The Breit frame is characterized by the vanishing energy component of the 
momentum of the exchanged photon. Both the photon momentum  
\begin{equation}
q=(0,0,0,-2xE),\,\,\,-q^2=Q^2=4x^2E^2\\
\end{equation}
and the proton momentum
\begin{equation}
P=E(1,0,0,1)
\end{equation}
are chosen along the $z$-direction.  
$x$ is the standard Bjorken scaling variable.
In the parton model, the incoming quark 
with momentum $p=xE(1,0,0,1)$ collides elastically with the virtual
boson and is scattered 
in the opposite direction (the 'current hemisphere')
with momentum  $p'=xE(1,0,0,-1)$;
therefore, $(k_T^{B}(p'))^2= Q^2$ in the limit of the quark parton model,
whereas $p_T^B(p')=0$.

The kinematics for dijet production is more complex: the momentum fraction
$\eta$ of the incoming parton must be larger than $x$ since 
$m_{jj}^2= Q^2(\eta/x-1)$ and, in general,
the jets have a nonvanishing transverse momentum
with respect to the $\gamma^*$-proton direction.
At LO, i.e. for massless jets, the 
relation between $k_T^B(j)$ and $p_T^B(j)=E_j\sin\theta_{jP}$ reads:
\begin{equation}
k_T^B(j) = p_T^B(j)\sqrt{\frac{2}{1+\cos\theta_{{\,\!jP}}}}
\end{equation}
Obviously, $k_T^B(j)>p_T^B$, and one can also show that
$\sum_j \,k_T^B(j) > Q$. Thus, $\sum_j \,k_T^B(j)$ is approximately 
given by the harder of the two scales $Q$ and $\sum_j \,p_T^B(j)$
\cite{mz3}.
For large dijet invariant masses (i.e. for ``true'' two jet kinematics)
one has $\eta >> x$, the dijet system will 
be strongly boosted in the proton direction 
and $k_T^B(j) \approx p_T^B(j)>>Q/2$, 
as long as very small scattering angles in 
the center of mass frame are avoided. 
For $m_{jj}\approx Q$ or smaller, on the other hand, 
$\sum_j \,k_T^B(j)\approx Q$ and, typically, both are 
considerably larger than $\sum_j \,p_T^B(j)<m_{jj}$ (in particular for
$Q>>m_{jj}$, which corresponds to the parton model limit).

Thus $\sum_j \,k_T^B(j)$ smoothly interpolates between the correct limiting
scale choices, it approaches $Q$ in the parton limit and it corresponds to 
the jet
transverse momentum when the photon virtuality becomes negligible.
It appears to be the ``natural'' scale for multi jet production in DIS.

In \cite{mz1} we found that the scale dependence of the dijet cross section
does not markedly improve in NLO for $\mu^2=\xi Q^2$.
This is also shown in Fig.~1 (dotted curves)
where  the dependence of the two-jet 
cross section  on the scale factor $\xi$ is shown.
We used the parton distribution functions set MRS D-$^\prime$
\protect\cite{mrs}
and employed the two loop formula for the strong coupling constant
both in the LO and NLO curves.
For scales related to  $\sum_i \,p_T^B(i)$ the uncertainty from the variation
of the renormalization and factorization scale is markedly reduced
compared to the LO predictions (dashed curves in Fig.~1).
Here $\xi$ is  defined via
\begin{equation}
 \mu_R^2 = \mu_F^2 = \xi\;(\sum_i \,p_T^B(i))^2\,. 
\label{xidef}
\end{equation}
%
%
The resulting $\xi$ dependence for
$ \mu_R^2 = \mu_F^2 = \xi\;(\sum_i \,k_T^B(i))^2$
is shown  as the solid lines in Fig.~1.
In this case, the NLO two-jet cross section is essentially independent on
$\xi$ for $\xi<2$.
Hence, the theoretical uncertainties due to the scale variation
are very small suggesting a precise
determination of $\alpha_s(<k_T^B>)$ for different $<k_T^B>$ bins,
where 
\begin{equation}
<k_T^B>=\frac{1}{2}\,\, (\sum_{j=1,2} \,k_T^B(j))
\label{akt}
\end{equation}
\begin{figure}[tp]
\vspace{5.5cm}
\begin{picture}(7,7)
\includegraphics{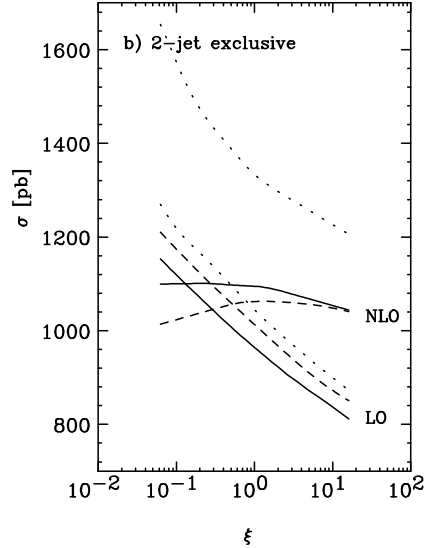}
\end{picture}
\caption{
Dependence of the two-jet exclusive cross 
section in the cone scheme 
on the  scale factor $\xi$.
The dashed curves are for $\mu_R^2=\mu_F^2=\xi\;(\sum_i\;p_T^B(i))^2$.
Choosing $(\sum_i\;k_T^B(i))^2$ as the basic scale yields the solid curves.
Choosing $Q^2$ as the basic scale yields the dotted curves.
Results are shown 
for the LO (lower curves) and NLO calculations.
}
\end{figure}

\begin{figure}[tp]
\vspace{5.5cm}
\begin{picture}(7,7)
\includegraphics{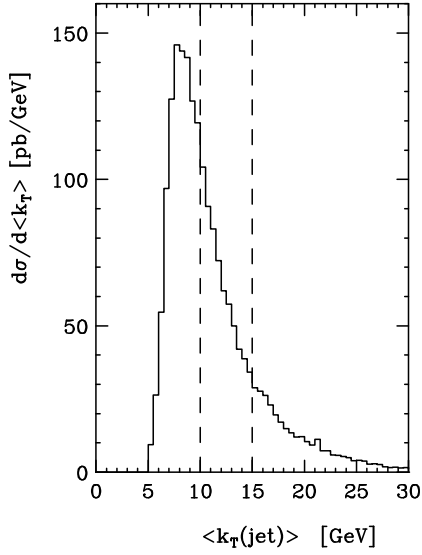}
\end{picture}
\caption{
$<k_T^B>$ distribution for the two-jet exclusive cross section.
The parameters are explained in the text.
}
\end{figure}
\begin{figure}[tp]
\vspace{5.5cm}
\begin{picture}(7,7)
\includegraphics{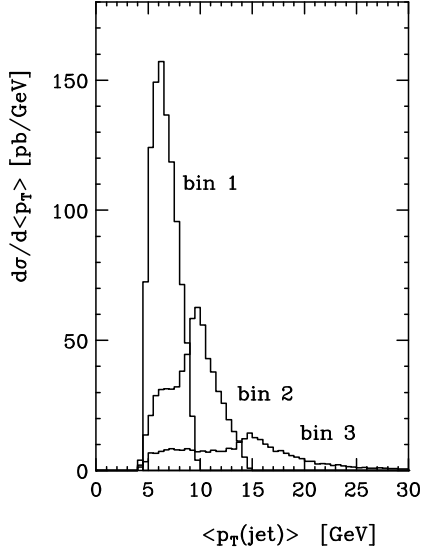}
\end{picture}
\caption{
$<p_T^B>$ distribution for the three $<k_T^B>$ bins shown in Fig.~1.
}
\end{figure}

\begin{figure}[tp]
\vspace{5.5cm}
\begin{picture}(7,7)
\includegraphics{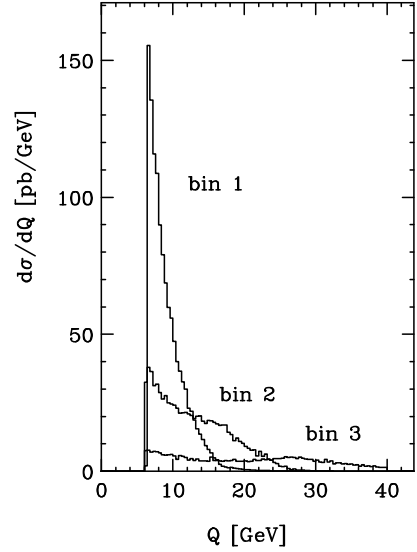}
\end{picture}
\caption{
$Q^2$ distribution  for the three $<k_T^B>$ bins shown in Fig.~1.
}
\end{figure}
Fig.~2 shows the $<k_T^B>$ distribution for the NLO exclusive dijet
cross section.
We used the parton distribution functions set GRV \cite{grv}
and  $\mu_R^2=\mu_F^2=1/4\;(\sum_i\;k_T^B(i))^2$.
In addition to the cuts imposed in Fig.~1,
we require $p_T^B>4$~GeV for each jet 
and $\sum_j k_T^B(j)> 10$ GeV.
We have divided the NLO cross section in the following three
$<k_T^B>$ bins.\\
bin 1: 5~GeV $<\,\,\,\,<k_T^B>\,\, \,\,<$ 10~GeV \\
bin 2: 10~GeV $<\,\,\,\,<k_T^B> \,\,\,\,<$ 15~GeV \\
bin 3: 15~GeV $<\,\,\,\,<k_T^B>$.\\
Table~1 shows the  corresponding NLO cross sections 
for these three bins for two different sets of parton distributions
and for different values for the scale factor $\xi$.
The theoretical uncertainties
for the
NLO dijet cross section from these numbers
are very small, in particular for the first two bins.
\begin{table}[t]
\caption{Jet cross sections in pb  for the three $<k_T^B>$ bins. 
}\label{table1}
\vspace{2mm}
\begin{tabular}{lccc}
        \hspace{0.8cm}
     &  \mbox{bin 1  }
     &  \mbox{bin 2  }
     &  \mbox{bin 3  } \\
\hline\\[-3mm]
\mbox{GRV}    $\xi=1/4 $ & 497    &  320  & 146    \\
\mbox{GRV}    $\xi=1/16$ & 504    &  306  & 134    \\
\mbox{GRV}    $\xi=1   $ & 488    &  322  & 151    \\
\mbox{MRSD-'} $\xi=1/4 $ & 487    &  322  & 148    \\
\end{tabular}
\end{table}
Fig.~3 and 4 show the $<p_T^B>$ and $Q^2$
 distribution for the NLO exclusive dijet
cross section for these three bins.
Whereas the $<p_T^B>$ and $Q^2$ distributions are fairly similar
to the $<k_T^B>$ distribution for the lowest bin, large differences
are found for the other bins. 
This partly  explains the rather different scale dependence observed
in Fig.1.

\section{Forward Jet Production in the Low $x$ Regime}
Deep inelastic scattering with a measured forward jet 
with relatively large momentum fraction $x_{jet}$
(in the proton direction) and $p_T^{2\,lab}(j)\approx Q^2$
is expected to provide 
sensitive information about the BFKL dynamics at low $x$
\cite{mueller,allen1}.
In this region there is not much phase space 
for DGLAP evolution with transverse momentum ordering,
whereas large effects are expected for BFKL evolution in $x$.
In particular, BFKL evolution is expected to substantially enhance cross
sections in the region $x<<x_{jet}$ \cite{mueller,allen1}.
In order to extract information on the $\ln(1/x)$
BFKL evolution, one needs to show that cross section results based on fixed 
order QCD with DGLAP evolution are not sufficient to describe the data. 
Clearly, next-to-leading order QCD corrections to the DGLAP predictions
are needed to make this comparison between experiment and theory.

In Table~\ref{table2} we show numerical results for the multi jet cross 
sections with (or without) a forward jet. 
The LO (NLO) results are based on 
the LO (NLO) parton distributions from GRV \cite{grv} together with
the one-loop (two-loop) formula 
for the strong coupling constant.
Kinematical cuts are imposed to closely model the H1 event 
selection\cite{deroeck}. More specifically, we require 
$Q^2>8~$GeV$^2$ , $x<0.004$,
$0.1 < y < 1$, an energy cut of $E(e^\prime)>11$~GeV on the scattered 
electron, and a cut on the pseudo-rapidity $\eta=-\ln\tan(\theta/2)$
of the scattered lepton of 
$ -2.868 < \eta(e^\prime)< -1.735$ 
(corresponding to $160^o < \theta(l^\prime) < 173.5^o$).
Jets are defined in the cone scheme (in the laboratory frame) with
$\Delta R = 1$ and $|\eta(j)|<3.5$.
We require a forward jet with $x_{jet}=p_z(j)/E_{P} > 0.05$,
$E(j)>25$ GeV, $0.5<p_T^2(j)/Q^2<4$,
and a cut on the pseudo-rapidity of 
$ 1.735< \eta(j)< 2.9$ 
(corresponding to $6.3^o < \theta(j) < 20^o$).
In addition all jets must have 
transverse momenta of at least  4 GeV in the lab frame
and 2 GeV in the Breit frame.

\begin{table}[t]
\caption{Cross sections for $n$-jet exclusive events
in DIS at HERA. See text for details.  
}\label{table2}
\vspace{2mm}
\begin{tabular}{lcc}
        \hspace{0.8cm}
     &  \mbox{with  }
     &  \mbox{without  } \\
     &  \mbox{forward jet}
     &  \mbox{forward jet}\\
\hline\\[-3mm]
\mbox{1 jet (LO)}   & 0    pb &  9026 pb       \\
\mbox{2 jet (LO)}   & 19.3 pb &  2219 pb    \\
\mbox{2 jet (NLO)}  & 68   pb &  2604 pb    \\
\mbox{3 jet (LO)}   & 30.1   pb & 450  pb    \\
\end{tabular}
\end{table}

The cross sections of Table~\ref{table2} demonstrate first of all that
the requirement of a forward jet with large longitudinal momentum fraction
($x_{jet}>0.05$) and restricted transverse momentum ($0.5<p_T^2(j)/Q^2<4$)
severely restricts the available phase space, in particular for low jet
multiplicities. The 1-jet exclusive cross section vanishes at LO, due to the
contradicting $x<0.004$ and $x_{jet}>0.05$ requirements. For $x<<x_{jet}$,
a high invariant mass hadronic system must be produced by the photon-parton
collision and this condition translates into 
\begin{eqnarray}
2E(j)m_T\;e^{-y} &\approx& \hat{s}_{\gamma,parton} \nonumber\\ 
                 &\approx& Q^2\left({x_{jet}\over x}-1\right) >> Q^2\; ,
\end{eqnarray}
where $m_T$ and $y$ are the transverse mass and rapidity of the 
partonic recoil system, respectively. Thus a recoil system with substantial
transverse momentum and/or invariant mass must be produced and this 
condition favors recoil systems composed out of at least two additional 
energetic partons. 

As a result 
one finds very large fixed order perturbative QCD corrections (compare
2 jet LO and NLO results with a forward jet in Table~\ref{table2}). 
In addition, the LO $({\cal{O}}(\alpha_s^2))$ 3-jet cross section 
is larger than the LO $({\cal{O}}(\alpha_s))$
2-jet cross section.
Thus, the forward jet  cross sections in Table~\ref{table2} are dominated
by the $({\cal{O}}(\alpha_s^2))$ matrix elements.
The effects
of BFKL evolution must be seen and isolated on top of these fixed order QCD
effects. We will analyze these effects in a subsequent publication.  

\section{Conclusions}

The calculation of NLO perturbative QCD corrections has received an enormous
boost with the advent of full NLO Monte Carlo programs 
\cite{giele1,jim}. For dijet production at HERA the NLO Monte Carlo
program MEPJET \cite{mz1} allows to study jet cross sections for arbitrary
jet algorithms. Internal jet structure, parton/hadron recombination effects, 
and the effects of arbitrary acceptance cuts can now be simulated at the full
${\cal O}(\alpha_s^2)$ level.
We found large NLO effects for some jet definition schemes (in particular the
$W$-scheme) and cone and $k_T$ schemes appear better suited for precision
QCD tests. 

The extraction of gluon distribution functions is now supported by
a fully versatile NLO program. Preliminary studies show that large NLO 
corrections are present in the Bjorken $x$ distribution for dijet events,
while these effects are mitigated in the reconstructed Feynman $x$ ($x_i$)
distribution, thus aiding the reliable extraction of $g(x_i,\mu_F^2)$.

For the study of BFKL evolution by considering events with a forward 
``Mueller''-jet very large  
QCD corrections are found at ${\cal O}(\alpha_s^2)$.
These fixed order effects form an important background to the observation
of BFKL evolution at HERA. They can now be studied systematically and for
arbitrary jet algorithms.

This research was supported by the University of Wisconsin 
Research Committee with funds granted by the Wisconsin Alumni Research 
Foundation and by the U.~S.~Department of Energy under Grant 
No.~DE-FG02-95ER40896. The work of E.~M. was supported in part  
by DFG Contract Ku 502/5-1.

\end{document}